\documentclass[aps,preprint]{revtex4}%
\usepackage{amsfonts}
\usepackage{amsmath}
\usepackage{amssymb}
\usepackage{graphicx}%
\newtheorem{theorem}{Theorem}

\newtheorem{corollary}[theorem]{Corollary}

\newenvironment{proof}[1][Proof]{\noindent\textbf{#1.} }{\ \rule{0.5em}{0.5em}}
\begin{document}
\title{Geometry of quantum correlations}
\author{Itamar Pitowsky}
\affiliation{Department \ of Philosophy, The Hebrew University, Mount Scopus, Jerusalem
91905, Israel.}
\email{itamarp@vms.huji.ac.il}
\published{\textit{Physical Review A }\textbf{77}, 062109 (2008).\textit{ }}

\begin{abstract}
Consider the set $\mathcal{Q}$ of quantum correlation vectors for two
observers, each with two possible binary measurements. Quadric (hyperbolic)
inequalities which are satisfied by every $q\in\mathcal{Q}$ are proved, and
equality holds on a two dimensional manifold consisting of the local boxes,
and all the quantum correlation vectors that maximally violate the Clauser,
Horne, Shimony, and Holt (CHSH)\ inequality. The quadric inequalities are
tightly related to CHSH, they are their iterated versions (equation 20).
Consequently, it is proved that $\mathcal{Q}$ is contained in a hyperbolic
cube whose axes lie along the non-local (Popescu, Rohrlich) boxes. As an
application, a tight constraint on the rate of local boxes that must be
present in every quantum correlation is derived. The inequalities allow
testing the validity of quantum mechanics on the basis of data available from
experiments which test the violation of CHSH. It is noted how these results
can be generalized to the case of $n$ sites, each with two possible binary measurements.

\end{abstract}
\maketitle

\section{Introduction}

The non-local character of quantum correlations is manifested by the violation
of Bell inequality \cite{1}, and more generally the Clauser, Horne, Shimony,
and Holt (CHSH) inequality \cite{2}. This property has become one of the
cornerstones of quantum information theory; beginning with Ekert's observation
\cite{3} that the violation of CHSH can be applied to protect the security of
key distribution, the number of publications on this subject is growing at a
fast rate. Still it is not completely clear why should quantum correlations
violate locality the way they do.

A fresh perspective on this problem was added by Popescu and Rohrlich
\cite{4}. They demonstrated that there are non-local correlations that do not
allow superluminal signaling, but nevertheless violate CHSH more strongly than
any quantum correlations (and therefore cannot be realized as far as present
day physics is concerned). The extreme form of these correlations became known
as PR-boxes. Despite their fictitious nature they shed new light on some
information theoretic problems. Thus, for example, quantum correlations
sometimes provide exponential gain in communication complexity over classical
correlations \cite{5}, while the availability of PR-boxes trivializes
communication complexity entirely \cite{6}, \cite{7}. Now we can ask a
complementary question: why is it that quantum correlations do not extend all
the way to the PR-boxes?

The relations between local correlations, quantum correlations, and the
PR-boxes have a geometric representation. Imagine a source of pairs of
particles, one goes to Alice and the other to Bob. Both Alice and Bob are
equipped with communication boxes, each box has two settings which will be
denoted by the index $i=1,2$ for Alice, and $j=1,2$ for Bob. In each run a
pair of particles is sent from the source, and Alice and Bob freely choose
their settings $i$ and $j$. When the particles arrive to the boxes an outcome
is registered in each box, which is either $+1$ or $-1$. Let $s_{ij}=\pm1$ be
the product of Alice's outcome and Bob's outcome. Repeat the runs many times
for the setting $ij$, denote the average by $p_{ij}$, and repeat this for all
four settings. The vector $p=(p_{11},p_{12},p_{21},p_{22})$ is called
correlation vector.

The \emph{local polytope }$\mathcal{L}$ is defined to be the subset in
$\mathbb{R}^{4}$ of all correlation vectors such that $p_{ij}=E(X_{i}Y_{j})$,
where $X_{i}$, $Y_{j}$ are real random variables on an arbitrary probability
space $(\Lambda,\Sigma,\mu)$ having values $\pm1$, and $E(X_{i}Y_{j})=\int
X_{i}(\lambda)Y_{j}(\lambda)d\mu(\lambda)$ are the expectations. $\mathcal{L}$
is the convex hull in $\mathbb{R}^{4}$ of the eight vertices,%

\begin{equation}%
\begin{array}
[c]{cccc}%
l_{1}=(1,1,1,1) & l_{2}=(1,1,-1,-1) & l_{3}=(1,-1,1,-1) & l_{4}=(1,-1,-1,1)\\
-l_{1}=(-1,-1,-1,-1) & -l_{2}=(-1,-1,1,1) & -l_{3}=(-1,1,-1,1) &
-l_{4}=(-1,1,1,-1)
\end{array}
. \label{1}%
\end{equation}
The facet inequalities are the eight trivial inequalities,%
\begin{equation}
-1\leq p_{ij}\leq1\quad i,j=1,2, \label{2}%
\end{equation}
and the eight Clauser, Horne, Shimony, Holt \ (CHSH) inequalities \cite{8},
\cite{9},
\begin{equation}
-1\leq\frac{1}{2}p_{11}+\frac{1}{2}p_{12}+\frac{1}{2}p_{21}+\frac{1}{2}%
p_{22}-p_{ij}\leq1\ \ \ \ i,j=1,2. \label{3}%
\end{equation}
The \emph{Popescu Rohrlich polytope }$\mathcal{P}$ \cite{10} is obtained by
adding eight more vertices, the PR-boxes, to those in (\ref{1}),
\begin{equation}%
\begin{array}
[c]{cccc}%
n_{1}=(-1,1,1,1) & n_{2}=(1,-1,1,1) & n_{3}=(1,1,-1,1) & n_{4}=(1,1,1,-1)\\
-n_{1}=(1,-1,-1,-1) & -n_{2}=(-1,1,-1,-1) & -n_{3}=(-1,-1,1,-1) &
-n_{4}=(-1,-1,-1,1)
\end{array}
, \label{4}%
\end{equation}
and the inequalities for $\mathcal{P}$ are just the trivial inequalities in
(\ref{2}).

As mentioned above, the vertices of $\mathcal{P}$ can be associated with the
the product of outputs of (real or hypothetical) communication boxes. The
eight vertices of $\mathcal{L}$ correspond to \emph{local boxes} that can
easily be realized. To see that think about the source as emitting pairs of
balls such that the two balls in each pair are of the same color, and the
colors are randomly distributed so that 50\% of the pairs are red and 50\%
black. We assume that at the outset, before any experiment is run, Alice and
Bob agree on the random variables $X_{i}$ and $Y_{j}$, but afterwords they
have no communication between them. Suppose that Alice's first setting, $i=1$,
is "$X_{1}=1$ if the ball is red, and $X_{1}=-1$ if it is black". To realize
the vertex $-l_{1}=(-1,-1,-1,-1)$ Alice choose $X_{2}=X_{1}$, and Bob chooses
$Y_{1}=Y_{2}=-X_{1}$, and in this case the outputs on both sides are perfectly
anti-correlated. \ To realize the vertex $l_{4}=(1,-1,-1,1)$ choose
$Y_{1}=X_{1}$ and $X_{2}=Y_{2}=-X_{1}$, in which case Alice and Bob outputs
are perfectly correlated in the second and third experiment, and perfectly
anti-correlated in the others.

The PR-boxes (\ref{4}) cannot be realized in a similar manner, as far as
present day physics is concerned. Take for example the vertex $n_{4}%
=(1,1,1,-1)$. \ For the first three set-ups $i,j=1,1$, or $1,2$, or $2,1$
Alice and Bob observe balls of the same color, and in the last setting
$i,j=2,2$ they detect different colors. There are no classical local random
variables $X_{i}$, $Y_{j}$ with the above properties, which can be chosen in
advance to yield these outcomes, nor are they quantum states and measurements
capable of producing it. However, all the boxes in $\mathcal{P}$, real or
imaginary, satisfy the important physical restriction of \emph{no signaling}.
This means that Bob cannot signal to Alice by changing his setting, say from
$j=1$ to $j=2$. In the above example all Alice detects are 50\% red balls and
50\% black balls, no matter what Bob is doing, and the same applies to Alice.

The outputs of quantum mechanical experiments lie in between the two
polytopes, there are quantum correlation vectors $q=(q_{11},q_{12}%
,q_{21},q_{22})$ such that $q\in\mathcal{P}\setminus\mathcal{L}$. Let $\rho$
be any quantum state defined on the tensor product of two Hilbert spaces,
$\mathbb{H=H}_{1}\otimes\mathbb{H}_{2}$. Suppose $A_{i}$, $i=1,2$, are
Hermitian operators on $\mathbb{H}_{1}$, and $B_{j}$, $j=1,2$, on
$\mathbb{H}_{2}$, such that their spectrum is in the interval $[-1,1]$. The
general quantum correlation vector has the form,
\begin{equation}
q_{ij}=tr(\rho(A_{i}\otimes B_{j})). \label{5}%
\end{equation}

Tsirelson \cite{11} proved that without loss of generality we can assume that
$\mathbb{H}_{1}\mathbb{=H}_{2}=\mathbb{C}^{2}$, where $\mathbb{C}$ is the
complex field; and for four directions (unit vectors) in physical space
$\mathbf{u}_{i}$, $\mathbf{v}_{j}$, $i,j=1,2,$ we can set $A_{i}%
=\sigma_{\mathbf{u}_{i}}$, and $B_{j}=\sigma_{\mathbf{v}_{j}}$, where the
$\sigma$'s are the spin operators in the corresponding directions. With this
representation consider a source of pairs in the state $\rho$ that emits the
particles towards Alice and Bob. For each run of the experiment Alice can
choose to measure either $\sigma_{\mathbf{u}_{1}}$ or $\sigma_{\mathbf{u}%
_{\mathbf{2}}}$ with possible outcomes $\pm1$, and Bob can choose between
$\sigma_{\mathbf{v}_{1}}$ and $\sigma_{\mathbf{v}_{\mathbf{2}}}$. The
correlation vector is then given by $q_{ij}=tr(\rho(\sigma_{\mathbf{u}_{i}%
}\otimes\sigma_{\mathbf{v}_{j}}))$.

Denote by $\mathcal{Q}$ the set of all vectors $q\in\mathbb{R}^{4}$ that have
this form, as we vary $\rho$ and the directions $\mathbf{u}_{i}$,
$\mathbf{v}_{j}$. The body $\mathcal{Q}$ is convex and satisfies
$\mathcal{L\subsetneqq Q\subsetneqq P}$ . Its structure has been described by
Tsirelson \cite{11}, and subsequently in different equivalent ways \cite{12},
\cite{13}, \cite{14}, \cite{15} the latter is the most compact representation
given by the inequalities%
\begin{equation}
\left\vert q_{11}q_{12}-q_{21}q_{22}\right\vert \leq\sqrt{1-q_{11}^{2}}%
\sqrt{1-q_{12}^{2}}+\sqrt{1-q_{21}^{2}}\sqrt{1-q_{22}^{2}}. \label{6}%
\end{equation}
The boundary $\partial\mathcal{Q}$ is a complicated $3$-dimensional algebraic
manifold. This mathematical description has been known for a while but its
physical significance is little understood. The purpose of this paper is to
further advance the analysis of the structure of $\mathcal{Q}$. Mathematically
I will demonstrate that $\mathcal{Q}$ is contained in a $4$-dimensional
hyperbolic cube, whose axes lie along the PR-boxes, and whose boundary is
given by quadric inequalities which are directly related to the CHSH
inequalities (\ref{3}), in fact they are iterated versions of CHSH (see
\ref{20} below, other quadric inequalities satisfied by all $q\in Q$ have been
previously derived in \cite{16}). Moreover, the intersection of the boundary
of the hyperbolic cube with $\partial\mathcal{Q}$ is a $2$-dimensional
sub-manifold of $\partial\mathcal{Q}$ corresponding to maximal quantum
violations of CHSH, as explained in theorem 1 below. The physical consequences
are examined subsequently, and include a calculation of the rate of local
boxes that must be present in every quantum correlation vector.

\section{Mathematical results}

\ The first thing to notice is that $\mathcal{P}$ is just the $4$-dimensional
unit cube, and $\mathcal{L}$ is the $4$-dimensional octahedron, so that they
are polar (dual) to each other. However, while $\mathcal{P}$ is presented in
its canonical form, the $4$-octahedron $\mathcal{L}$ is rotated from its
canonical representation, which is just the convex hull of%
\begin{equation}%
\begin{array}
[c]{cccc}%
e_{1}=(1,0,0,0) & e_{2}=(0,1,0,0) & e_{3}=(0,0,1,0) & e_{4}=(0,0,0,1)\\
-e_{1}=(-1,0,0,0) & -e_{2}=(0,-1,0,0) & -e_{3}=(0,0,-1,0) & -e_{4}=(0,0,0,-1)
\end{array}
. \label{7}%
\end{equation}
The matrix that transforms the vertices of $\mathcal{L}$ \ in (\ref{1}) to the
respective vertices of the canonical form in (\ref{7}) is%
\begin{equation}
H=\frac{1}{4}\left(
\begin{array}
[c]{cccc}%
1 & 1 & 1 & 1\\
1 & 1 & -1 & -1\\
1 & -1 & 1 & -1\\
1 & -1 & -1 & 1
\end{array}
\right)  , \label{8}%
\end{equation}
so that $2H$ is an orthogonal self adjoint Hadamard matrix. We shall denote
the $4$-octahedron in the canonical form by $H\mathcal{L}$. The facet
inequalities of $H\mathcal{L}$ have a particularly simple form. If
$r=(r_{11},r_{12},r_{21},r_{22})\in H\mathcal{L}$ then the facet inequalities
are%
\begin{equation}
\sum_{i,j=1,2}\left\vert r_{ij}\right\vert \leq1. \label{9}%
\end{equation}

Denote by $\partial\mathcal{L}_{ij}$ the facet of $\mathcal{L}$ that
corresponds to an equality on the right hand side of \ (\ref{3}) for $i$, $j$.
For example for $i=j=2$,
\begin{equation}
\partial\mathcal{L}_{22}=co\{l_{1},l_{2},l_{3},-l_{4}\}, \label{10}%
\end{equation}
Where $co$ stands for the convex hull. This facet is transformed by $H$ to,%
\begin{equation}
H(\partial\mathcal{L}_{22})=co\{e_{1},e_{2},e_{3},-e_{4}\}. \label{11}%
\end{equation}
In general, any non trivial facet (\ref{3}) of $\mathcal{L}$ is transformed by
$H$ to a convex hull of four vertices with an odd number of negated $e_{i}$'s,
and every trivial facet (\ref{2}) moves by $H$ to the convex hull of an even
number (including zero) of negated $e_{i}$'s.

Another important feature is that all the PR-boxes in (\ref{4}) are
eigenvectors of $2H$, with $\pm n_{1}$ corresponding to the eigenvalue $-1$,
and the others corresponding to the eigenvalue $+1$. Also, all PR-boxes are
either opposite each other or orthogonal to each other in $\mathbb{R}^{4}$.
Hence the quadric form,%
\begin{equation}
q^{t}Hq=\frac{1}{4}(q_{11}+q_{12}+q_{21}-q_{22})^{2}+(q_{11}q_{22}%
-q_{12}q_{21}), \label{12}%
\end{equation}
defines "Minkowskian metric" in $\mathbb{R}^{4}$, with the axis along $\pm
n_{1}$ playing the role of "time", and the other PR-boxes the "space" axes.
The surface $q^{t}Hq=1$ is thus a hyperboloid. We have

\begin{theorem}
If $q\in Q$ then $q^{t}Hq\leq1$, and equality holds on a two dimensional
submanifold of $\partial\mathcal{Q}$ which includes all the local boxes, and
the a subset of $\partial\mathcal{Q}$ which maximally violate the CHSH
inequality. We also have $q^{t}Hq\geq-1$ for all $q\in\mathcal{Q}$.
\end{theorem}

\begin{proof}
We shall use the following characterization due to Tsirelson \cite{11}. If
$q\in\mathcal{Q}$ there are unit vectors $\mathbf{x}_{1},\mathbf{x}%
_{2},\mathbf{y}_{1},\mathbf{y}_{2}\in\mathbb{R}^{4}$ such that $q_{ij}%
=\mathbf{x}_{i}\cdot\mathbf{y}_{j}$ for $i,j=1,2$. Moreover, if $q\in
\partial\mathcal{Q}$ then the $\mathbf{x}_{i}$,$\mathbf{y}_{j}$'s are in the
same plane. Put%
\begin{equation}
\mathbf{a=}\frac{1}{2}(\mathbf{x}_{1}+\mathbf{x}_{2}),\quad\mathbf{a}^{\bot
}\mathbf{=}\frac{1}{2}(\mathbf{x}_{1}-\mathbf{x}_{2}),\quad\mathbf{b=}\frac
{1}{2}(\mathbf{y}_{1}+\mathbf{y}_{2}),\quad\mathbf{b}^{\bot}\mathbf{=}\frac
{1}{2}(\mathbf{y}_{1}-\mathbf{y}_{2}), \label{13}%
\end{equation}
then $\mathbf{a}$ and $\mathbf{a}^{\bot}$ are orthogonal to each other,
$\mathbf{b}$ and $\mathbf{b}^{\bot}$ are orthogonal, $\left\Vert
\mathbf{a}\right\Vert ^{2}+\left\Vert \mathbf{a}^{\bot}\right\Vert ^{2}=1$ and
$\left\Vert \mathbf{b}\right\Vert ^{2}+\left\Vert \mathbf{b}^{\bot}\right\Vert
^{2}=1$ where $\left\Vert .\right\Vert $ is the Euclidean norm in
$\mathbb{R}^{4}$. Then a straightforward calculation shows%
\begin{equation}
(Hq)_{11}=\mathbf{a\cdot b},\mathbf{\quad(}Hq\mathbf{)}_{12}=\mathbf{a}^{\bot
}\cdot\mathbf{b},\mathbf{\quad}(Hq)_{21}=\mathbf{a\cdot b}^{\bot}%
,\quad(Hq)_{22}=\mathbf{a}^{\bot}\cdot\mathbf{b}^{\bot}. \label{14}%
\end{equation}

Put $\left\Vert \mathbf{a}\right\Vert =\cos\alpha$, and $\left\Vert
\mathbf{b}\right\Vert =\cos\beta$. Assume that $q\in\partial\mathcal{Q}$ and
the $\mathbf{x}_{i}$, $\mathbf{y}_{j}$'s are in the same plane and let
$\theta$ be the angle between $\mathbf{a}$ and $\mathbf{b}$. If $q$ is in the
part of $\partial\mathcal{Q}$ just above the facet $\partial\mathcal{L}_{22}$
in (\ref{10}), we deduce from (\ref{11}) that $(Hq)_{11},(Hq)_{12}%
,(Hq)_{21}\geq0$ and $(Hq)_{22}\leq0$. Using (\ref{9}) we can calculate the
value of the CHSH expression,%
\begin{equation}
CHSH=\sum_{i,j=1,2}\left\vert (Hq)_{ij}\right\vert =\cos(\alpha-\beta
)\cos\theta+\sin(\alpha+\beta)\sin\theta. \label{15}%
\end{equation}
Suppose that we have fixed the lengths $\left\Vert \mathbf{a}\right\Vert
=\cos\alpha$, and $\left\Vert \mathbf{b}\right\Vert =\cos\beta$, then the
maximum on the right hand side of (\ref{15}) is obtained for $\theta$ which
satisfies%
\begin{equation}
\tan\theta=\frac{\sin(\alpha+\beta)}{\cos(\alpha-\beta)}. \label{16}%
\end{equation}
The value of the CHSH for this choice is,%
\begin{equation}
\max CHSH=\underset{\theta}{\max}\sum_{i,j=1,2}\left\vert (Hq)_{ij}\right\vert
=\sqrt{\cos^{2}(\alpha-\beta)+\sin^{2}(\alpha+\beta)}, \label{17}%
\end{equation}
with the absolute maximum $\sqrt{2}$ (the Tsirelson bound) obtained when we
take $\alpha=\beta=\frac{\pi}{4}$, (and $\theta=\frac{\pi}{4}$).

The matrix $2H$ is both self adjoint and orthogonal and therefore we have
$H^{2}=\frac{1}{4}I$. Substituting the values from (\ref{14}) to (\ref{12}) we
get,
\begin{equation}
q^{t}Hq=4(Hq)^{t}H(Hq)=[\cos(\alpha-\beta)\cos\theta+\sin(\alpha+\beta
)\sin\theta]^{2}-\sin(2\alpha)\sin(2\beta). \label{18}%
\end{equation}
Again, suppose that the lengths $\left\Vert \mathbf{a}\right\Vert =\cos\alpha
$, and $\left\Vert \mathbf{b}\right\Vert =\cos\beta$ are fixed, then for
$q\in\partial\mathcal{Q}$ above the facet $\partial\mathcal{L}_{22}$ the
maximum value of $q^{t}Hq$ is obtained at $\theta$ in (\ref{16}) and it is,%
\begin{equation}
\max(q^{t}Hq)=\cos^{2}(\alpha-\beta)+\sin^{2}(\alpha+\beta)-\sin(2\alpha
)\sin(2\beta)=1 \label{19}%
\end{equation}

It is straightforward to check that $q^{t}Hq=1$ for all the local boxes in
(\ref{1}). The quantum correlation vectors at which we obtain the absolute
extrema of CHSH are $\pm\frac{1}{\sqrt{2}}n_{k}$, where the $\pm n_{k}$ are
the PR-boxes (\ref{4}). Recall that the PR-boxes $n_{2}$, $n_{3}$, $n_{4}$ in
(\ref{4}) are also eigenvectors of $2H$ with eigenvalue $+1$ and therefore we
have $(\frac{1}{\sqrt{2}}n_{k})^{t}H(\frac{1}{\sqrt{2}}n_{k})=1$ for
$k=2,3,4$. Hence, the above argument can be repeated with regard to the part
of $\partial\mathcal{Q}$ above $\partial\mathcal{L}_{12}$ and above
$\partial\mathcal{L}_{21}$. Since $n_{1}$ is an eigenvector of $2H$ with
eigenvalue $-1$ we have $(\frac{1}{\sqrt{2}}n_{1})^{t}H(\frac{1}{\sqrt{2}%
}n_{1})=-1$, and the correlation vector $\frac{1}{\sqrt{2}}n_{1}$ does not lie
on the surface of the hyperboloid $q^{t}Hq=1$, nor does the part of
$\partial\mathcal{Q}$ above $\partial\mathcal{L}_{11}$; however we have
$q^{t}Hq\geq-1$ for all $q\in\mathcal{Q}$.
\end{proof}

\begin{corollary}
\bigskip The iterated CHSH: for all $q\in\mathcal{Q}$ we have%
\begin{align}
-1  &  \leq-\frac{1}{8}(-q_{11}+q_{12}+q_{21}+q_{22})^{2}+\frac{1}{8}%
(q_{11}-q_{12}+q_{21}+q_{22})^{2}+\smallskip\label{20}\\
&  \frac{1}{8}(q_{11}+q_{12}-q_{21}+q_{22})^{2}+\frac{1}{8}(q_{11}%
+q_{12}+q_{21}-q_{22})^{2}\leq1.\nonumber
\end{align}
and, by symmetry, another three inequalities of the same form, each with one
of the components of (\ref{20}) having a minus sign.
\end{corollary}

\begin{proof}
Since the $n_{k}$'s are orthogonal in pairs, we can write each correlation
vector $q=(q_{11},q_{12},q_{21},q_{22})$ in terms of the orthogonal basis
$\{n_{i}\}$. Since $Hn_{1}=-\frac{1}{2}n_{1}$ and $Hn_{k}=\frac{1}{2}n_{k}$
for $k=2,3,4$, this yields $q^{t}Hq=-\frac{1}{8}(n_{1}^{t}q)^{2}+\frac{1}%
{8}(n_{2}^{t}q)^{2}+\frac{1}{8}(n_{3}^{t}q)^{2}+\frac{1}{8}(n_{4}^{t}q)^{2}$,
and from theorem 1 we get (\ref{20}). From symmetry it is obvious that we can
choose any of the PR-boxes $\pm n_{k}$ to play the role of the "time"
(eigenvalue $=-1$) axis, and the other three the "space" axes, simply by
replacing the Hadamard matrix $2H$ by another. In this way we can get four
hyperboloids and $\mathcal{Q}$ is contained in their intersection, each yields
another inequality of the form (\ref{20}).
\end{proof}

\section{Physical consequences}

We can easily derive the experimental arrangements which will give rise to the
extrema (\ref{17}, \ref{19}). Using the fact that $H^{2}=\frac{1}{4}I$ we can
invert the relations in (\ref{14}), and represent $q_{ij}$ in terms of the
parameters $\alpha,\beta,\theta$%
\begin{equation}
q_{11}=\cos(\alpha+\beta-\theta),\ q_{12}=\cos(\alpha-\beta-\theta
),\ q_{21}=\cos(\alpha-\beta+\theta),\ q_{22}=\cos(\alpha+\beta+\theta)
\label{21}%
\end{equation}
with $\theta$ given by (\ref{16}). From these values the angles between the
directions $\mathbf{u}_{i}$ and $\mathbf{v}_{j}$ in the measurement of
$\sigma_{\mathbf{u}_{i}}\otimes\sigma_{\mathbf{v}_{j}}$ can be derived.

More generally, we can formulate the iterated CHSH is in term of the
observables $A_{i}$ and $B_{j}$ in (\ref{5}), denote for $i,j=1,2$,%
\begin{equation}
C_{ij}=\frac{1}{2}A_{1}\otimes B_{1}+\frac{1}{2}A_{1}\otimes B_{2}+\frac{1}%
{2}A_{2}\otimes B_{1}+\frac{1}{2}A_{2}\otimes B_{2}-A_{i}\otimes B_{j}.
\label{22}%
\end{equation}
Then the general iterated CHSH reads%
\begin{equation}
-1\leq\frac{1}{2}\left\vert tr(\rho C_{11})\right\vert ^{2}+\frac{1}%
{2}\left\vert tr(\rho C_{12})\right\vert ^{2}+\frac{1}{2}\left\vert tr(\rho
C_{21})\right\vert ^{2}+\frac{1}{2}\left\vert tr(\rho C_{22})\right\vert
^{2}-\left\vert tr(\rho C_{ij})\right\vert ^{2}\leq1, \label{23}%
\end{equation}
for $i,j=1,2$ and any state $\rho$ on $\mathbb{H}_{1}\otimes\mathbb{H}_{2}$.
Hence, the numbers $\left\vert tr(\rho C_{ij})\right\vert ^{2}$ satisfy all
the CHSH\ inequalities (\ref{3}), however, they do not necessarily satisfy the
trivial inequalities (\ref{2}). We can use this inequality to test the
validity of quantum mechanics, using the data that has already been collected
in many experiments that test the violation of CHSH. By quantum mechanics the
same data must satisfy the iterated CHSH, and the inequality is tight.

Perhaps the easiest way to grasp the interpretation these inequalities is in
terms of \emph{non-local deterministic hidden variable theories} such as
Bohm's (see e.g., \cite{17}) . Given the value of the hidden variable
$\lambda$ (in Bohm's theory, the exact initial positions of the particles of
an EPR pair) we can predict at the outset the outcomes of each of the four
measurements of $\sigma_{\mathbf{u}_{i}}\otimes\sigma_{\mathbf{v}_{j}}$,
$i,j=1,2$. Suppose that we want to recover the quantum correlation vector
$q=(q_{11},q_{12},q_{21},q_{22})$ that violates the CHSH inequality. We sample
at random the hidden variables $\lambda$ according to the measure $\mu$ on the
space of hidden variables (in Bohm's theory, initial values of the positions
of the particles according to the distribution $\left\vert \psi\right\vert
^{2}$ at time $0$, where $\psi$ is the full quantum state). For each value of
the hidden variable we calculate the deterministic outcomes of all four
experiments, the result is a $\pm1$ four-dimensional vector. Finally, to get
$q$, we take the average of the vectors. To obtain a result that violates CHSH
some of the $\pm1$ vectors in the sample must be PR-boxes, but how many? In
other words, what is the minimal frequency with which a PR-box should appear
in the hidden variable sample that yields the correlation vector $q$? (A
similar problem is considered in \cite{18}, \cite{19}).

Assume that $q\in Q$ is above the facet $\partial\mathcal{L}_{22}$ of
$\mathcal{L}$, given in (\ref{10}). In this case we can represent $q$ as a
convex combination%
\begin{equation}
q=\eta_{1}l_{1}+\eta_{2}l_{2}+\eta_{3}l_{3}-\eta_{4}l_{4}+\eta n_{4},\quad
\eta_{i},\eta\geq0,\quad\eta+%
{\displaystyle\sum}
\eta_{i}=1. \label{24}%
\end{equation}
The $l$'s are the local boxes in $\partial\mathcal{L}_{22}$ and $n_{4}$ is the
PR-box above $\partial\mathcal{L}_{22}$. Calculating $\eta$, the coefficient
of the PR box, we get,%
\begin{equation}
\eta=\frac{1}{2}(q_{11}+q_{12}+q_{21}-q_{22})-1\leq\sqrt{2}-1, \label{25}%
\end{equation}
and this is the minimal rate of the PR-box $n_{4}$ in the average (\ref{24}).
This result has an information theoretic formulation: Suppose that Alice and
Bob prepare a key using BB84, then $\eta=p_{NL}$ is the minimal rate with
which Eve should prepare and send a PR-box if she is to deceive Alice and Bob
that nobody listens on their line \cite{20}. Somewhat more mysteriously it is
also related to the critical security criteria of BB84 against symmetric
individual attacks \cite{21}.

We can also formulate the limitation on quantum correlations in terms of the
coefficients $\eta_{i}$ in (\ref{24}). Again, if we consider $q\in Q$ above
the facet $\partial\mathcal{L}_{22}$, the iterated CHSH inequality in
(\ref{12}, \ref{20}, \ref{23}) is equivalent to the formula%
\begin{equation}
\eta_{1}+\eta_{2}+\eta_{3}+\eta_{4}\geq1-2\sqrt{\eta_{1}\eta_{4}+\eta_{2}%
\eta_{3}}, \label{26}%
\end{equation}
with equality on the set of maximally violating quantum correlations described
in the proof of theorem 1. The interesting aspect about this inequality is
that it involves only the rates of the classical local boxes in our
hypothetical ensemble. This inequality bounds the rates of local boxes that
must be present in any quantum correlation vector $q$. In the symmetric case
when all the $\eta_{i}$'s are equal, $\eta_{i}=\eta_{0}$ we have%
\begin{equation}
\eta_{0}\geq\frac{1}{2}\left(  1-\frac{1}{\sqrt{2}}\right)  . \label{27}%
\end{equation}
and the total frequency of classical boxes that should be used to recover $q$
is $4\eta_{0}\geq2-\sqrt{2}$. The number on the right in (\ref{27}) is also
the critical value of the quantum bit error rate above which BB84 becomes
insecure against individual symmetric attacks \cite{21}.

It seems that these results can be readily generalized to the case of $n$
particles, and two binary traceless measurements on each. Werner and Wolf
\ \cite{13} established that the local correlation vectors (of dimension
$2^{n}$) form a polytope, with $2^{2^{n}}$ facet inequalities, all
generalizations of CHSH. The polytope is a $2^{n}$-dimensional octahedron. A
Hadamard matrix (with a suitable normalization) will transform this polytope
to its canonical position relative to its polar, the unit $2^{n}$-dimensional
cube. The Tsirelson boundary also has a detailed description in this case, and
it seems to me that the formulation and proof of theorem 1 can be repeated.

\begin{acknowledgments}
I would like to thank Nicolas Gisin and Daniel Rohrlich for their comments.
This research is supported by the Israel Science Foundation, grant number 744/07.
\end{acknowledgments}

\bigskip\

\end{document}